\RequirePackage{lineno}
\documentclass[apj]{emulateapj}

\usepackage{graphicx}
\usepackage{apjfonts}
\usepackage{amsmath}
\usepackage{color}
\usepackage{hyperref}

\newcommand{\psr}{PSR~J0737$-$3039A}
\newcommand{\whts}{41}
\newcommand{\whtssigma}{5.4}
\newcommand{\cutoffsigma}{4}
\newcommand{\g}{$\gamma$}

\shorttitle{PULSED GAMMA RAYS FROM \psr}
\shortauthors{\sc Guillemot et al.}

\begin{document}

\title{\emph{Fermi} LAT pulsed detection of \psr{} in the double pulsar system}

\author{
L.~Guillemot\altaffilmark{1},
M.~Kramer\altaffilmark{1,2},
T.~J.~Johnson\altaffilmark{3,9},
H.~A.~Craig\altaffilmark{4},
R.~W.~Romani\altaffilmark{4},
C.~Venter\altaffilmark{5},
A.~K.~Harding\altaffilmark{6},
R.~D.~Ferdman\altaffilmark{2,7},
I.~H.~Stairs\altaffilmark{8},
and M.~Kerr\altaffilmark{4}
}
\altaffiltext{1}{Max-Planck-Institut f\"ur Radioastronomie, Auf dem H\"ugel 69, D-53121 Bonn, Germany; guillemo@mpifr-bonn.mpg.de}
\altaffiltext{2}{Jodrell Bank Centre for Astrophysics, School of Physics and Astronomy, The University of Manchester, M13 9PL, UK}
\altaffiltext{3}{National Research Council Research Associate, National Academy of Sciences, Washington, DC 20001, USA}
\altaffiltext{4}{W. W. Hansen Experimental Physics Laboratory, Kavli Institute for Particle Astrophysics and Cosmology, Department of Physics and SLAC National Accelerator Laboratory, Stanford University, Stanford, CA 94305, USA}
\altaffiltext{5}{Centre for Space Research, North-West University, Potchefstroom Campus, Private Bag X6001, 2520 Potchefstroom, South Africa}
\altaffiltext{6}{NASA Goddard Space Flight Center, Greenbelt, MD 20771, USA}
\altaffiltext{7}{Department of Physics, Rutherford Physics Building, McGill University, 3600 University Street, Montreal, Quebec, H3A 2T8, Canada}
\altaffiltext{8}{Department of Physics and Astronomy, University of British Columbia, Vancouver, BC V6T 1Z1, Canada}
\altaffiltext{9}{Resident at Naval Research Laboratory, Washington, DC 20375, USA.}

\begin{abstract}

We report the \emph{Fermi} Large Area Telescope discovery of \g-ray pulsations from the 22.7 ms pulsar A in the double pulsar system J0737$-$3039A/B. This is the first mildly recycled millisecond pulsar (MSP) detected in the GeV domain. The 2.7 s companion object PSR~J0737$-$3039B is not detected in \g{} rays. PSR~J0737$-$3039A is a faint \g-ray emitter, so that its spectral properties are only weakly constrained; however, its measured efficiency is typical of other MSPs. The two peaks of the \g-ray light curve are separated by roughly half a rotation and are well offset from the radio and X-ray emission, suggesting that the GeV radiation originates in a distinct part of the magnetosphere from the other types of emission. From the modeling of the radio and the \g-ray emission profiles and the analysis of radio polarization data, we constrain the magnetic inclination $\alpha$ and the viewing angle $\zeta$ to be close to 90$^\circ$, which is consistent with independent studies of the radio emission from \psr{}. A small misalignment angle between the pulsar's spin axis and the system's orbital axis is therefore favored, supporting the hypothesis that pulsar B was formed in a nearly symmetric supernova explosion as has been discussed in the literature already. 

\end{abstract}


\section{Introduction}
\label{sec:introduction}

Pulsed GeV \g-ray emission from more than 100 pulsars\footnote{A list of \g-ray pulsar detections is available at\\ https://confluence.slac.stanford.edu/display/GLAMCOG/Public+List+of+LAT-Detected+Gamma-Ray+Pulsars.}  has been observed by the Large Area Telescope \citep[LAT;][]{FermiLAT} aboard the \emph{Fermi} satellite, launched in 2008 June \citep[see The Second Fermi Large Area Catalog of Gamma-Ray Pulsars;][hereafter 2PC]{Fermi2PC}. The current population of \g-ray pulsars includes objects known from independent radio or X-ray observations and detected in \g{} rays by folding the \emph{Fermi} LAT data at the known rotational periods \citep[e.g.,][]{Fermi8MSPs,FermiJ2043,Espinoza2013}, and \g-ray pulsars found through direct blind searches of the LAT data \citep{FermiBlindSearch,Pletsch9PSRs} or radio searches for pulsars in unassociated \g-ray sources \citep[see][and references therein]{Ray2012}. These pulsars are energetic (spin-down luminosities $\dot E = 4 \pi^2 I \dot P / P^3 > 10^{33}$ erg s$^{-1}$, where $P$ is the spin period, $\dot P$ is its first time derivative, and $I$ denotes the moment of inertia, assumed to be $10^{45}$ g cm$^2$ in this work) and are typically nearby. They therefore have large values of the ``spin-down flux'' $\dot E / d^2$ and of the heuristic detectability metric $\sqrt{\dot E} / d^2$ \citep{FermiPSRCatalog}.

The double pulsar system PSR~J0737$-$3039A/B \citep{Burgay2003,Lyne2004} consists of two radio-emitting neutron stars in a tight 2.4 hr orbit. Radio timing observations of the two pulsars provide high-precision tests of strong field gravity \citep{Kramer2006}. The 2.7 s pulsar J0737$-$3039B has a low spin-down luminosity $\dot E \sim 1.7 \times 10^{30}$ erg s$^{-1}$, three orders of magnitude smaller than that of the least energetic \g-ray pulsar currently known, making this pulsar unlikely to be detectable by the LAT. On the other hand, the higher $\dot E$ of $\sim 5.94 \times 10^{33}$ erg s$^{-1}$---for which the Shklovskii effect \citep{Shklovskii1970} is negligible due to the system's low transverse velocity---and the modest parallax distance of $1150_{-160}^{+220}$ pc measured with very long baseline interferometry observations \citep{Deller2009} make the 22.7 ms pulsar J0737$-$3039A a credible candidate for a detection in \g{} rays.

Searching for high-energy pulsations from PSRs J0737$-$3039A and B allows exploration of regions of the $P-\dot P$ diagram that are devoid of \g-ray pulsars \citep{FermiPSRCatalog}: mildly recycled millisecond pulsars (MSPs) for the former (mildly recycled pulsars being pulsars that were only partially spun-up by accretion of matter from a companion star), and slowly-rotating normal pulsars for the latter. Increasing the variety of \g-ray pulsar types helps understand the phenomenology of high-energy emission from these stars, and how emission properties evolve with, e.g., $\dot E$ or age (with the caveat that the evolution could be masked by effects arising from the geometry). Additionally, the modeling of pulse profiles at different wavelengths can yield constraints on the geometry of emission, in particular, the inclination angle of the magnetic axis ($\alpha$) and the observer viewing angle ($\zeta$) with respect to the spin axis. Constraints on $\alpha$ and $\zeta$ can also be determined by studying the radio polarization. Since the inclination of the binary system with respect to the line-of-sight is accurately known, $\zeta$ directly measures the misalignment between the pulsar's spin axis and the orbital angular momentum, a quantity which gives important clues about the formation of the binary system \citep[for a discussion, see, e.g.,][]{Ferdman2008}.

We here report the discovery of \g-ray pulsations from the 22.7 ms pulsar J0737$-$3039A, in 43 months of data recorded by the \emph{Fermi} LAT. In Section \ref{sec:gammaanalysis} we describe the analysis of the LAT data and the resulting light curve and spectrum for \psr{}.  In Sections \ref{sec:lcmodeling} and \ref{sec:pol}, we model the offset radio and \g-ray light curves, as well as the radio polarization, under various emission models, allowing us to constrain the pulsar's geometrical characteristics.  We conclude with a brief discussion on the implication of these results.

\section{LAT analysis}
\label{sec:gammaanalysis}

We analyzed \emph{Fermi} LAT events recorded between 2008 August 4 and 2012 March 7, belonging to the ``Source'' class under the P7\_V6 instrument response functions, and with zenith angles smaller than 100$^\circ$. We rejected events recorded when the rocking angle of the telescope exceeded 52$^\circ$, when the instrument was not operating in the science observations mode or when the data quality flag was not set as good. Analyses of the LAT data were carried out using the \emph{Fermi} Science Tools\footnote{http://fermi.gsfc.nasa.gov/ssc/data/analysis/scitools/overview.html} (STs) v9r28p0. The data were phase-folded with the \emph{Fermi} plug-in \citep{Ray2011} distributed with the \textsc{Tempo2} pulsar timing package \citep{tempo2}, and the ephemerides for PSRs J0737$-$3039A and B published in \citet{Kramer2006}. Since 2008 March PSR J0737$-$3039B has been invisible in the radio domain \citep{Perera2010} because of the precession of its spin axis \citep{Breton2008} causing its radio beam to miss the Earth, precluding timing measurements contemporaneous with the \emph{Fermi} mission. Additionally, changes in the radio profile of PSR~J0737$-$3039B prevented \citet{Kramer2006} from constructing a coherent timing model for this pulsar. Our ability to accurately fold the \emph{Fermi} LAT data using this ephemeris is therefore limited. Nevertheless, the very low spin-down luminosity of this pulsar makes it unlikely to be detected in \g{} rays.

From initial pulsation searches using standard data selection cuts we found no evidence of \g-ray pulsations from PSR~J0737$-$3039B. However, selecting events found within $0\fdg5$ from \psr{} and with energies above 0.2 GeV we obtained a value for the bin-independent \emph{H}-test parameter \citep{deJager2010} of 21.3, corresponding to a pulsation significance of $\sim 3.7\sigma$. The \g-ray pulse profile of PSR~J0737$-$3039A comprises two peaks, which is reminiscent of known \g-ray pulsar light curves \citep[e.g.,][]{FermiPSRCatalog}, with no evidence for emission in the $[0 ; 0.2] \cup [0.5 ; 0.8]$ phase interval.

\begin{deluxetable}{lc}
\tablewidth{0.95 \columnwidth}
\tablecaption{Main Properties and \g-Ray Parameters of \psr{} \label{tab:params}}
\tablecolumns{2}
\tablehead
{
\colhead{Parameter} &
\colhead{Value}
}
\startdata
Rotational period, $P$ (ms) \dotfill & $22.7$ \\
Period derivative, $\dot P$ ($10^{-18}$) \dotfill & $1.76$ \\
Spin-down luminosity, $\dot E$ ($10^{33}$ erg s$^{-1}$) \dotfill & $5.94$ \\
Magnetic field at the light cylinder, $B_{\rm LC}$ ($10^3$ G) \dotfill & $4.97$ \\
Distance, $d$ (pc) \dotfill & $1150^{+220}_{-150}$ \\
\hline
First peak position, $\Phi_1$ \dotfill & $0.43 \pm 0.01$ \\
First peak full width at half-maximum, FWHM$_1$ \dotfill & $0.03 \pm 0.02$ \\
Distance from the closest radio peak maximum, $\delta_1$ \dotfill & $0.20 \pm 0.01$ \\
Second peak position, $\Phi_2$ \dotfill & $0.93 \pm 0.01$ \\
Second peak full width at half-maximum, FWHM$_2$ \dotfill & $0.02_{-0.02}^{+0.03}$ \\
Distance from the closest radio peak maximum, $\delta_2$ \dotfill & $0.13 \pm 0.01$ \\
\g-ray peak separation, $\Delta = \Phi_2 - \Phi_1$ & $0.49 \pm 0.01$ \\
\hline
Photon index, $\Gamma$ \dotfill & $< 1.3$ \\
Cutoff energy, $E_c$ (GeV) \dotfill  & $ 0.4 \pm 0.4$ $(<0.9)$  \\
Photon flux, $F$ ($\geq 0.1$ GeV) ($10^{-9}$ photons cm$^{-2}$ s$^{-1}$)\dotfill & $ 6 \pm 3$ $(<9)$ \\
Energy flux, $G$ ($\geq 0.1$ GeV) ($10^{-12}$ erg cm$^{-2}$ s$^{-1}$) \dotfill & $ 4 \pm 1$ $(<5)$\\
Luminosity, $L_\gamma / f_\Omega = 4 \pi G d^2$ ($10^{32}$ erg s$^{-1}$) \dotfill & $ 6_{-2}^{+3}$ $(<8)$ \\
Efficiency, $\eta / f_\Omega = 4 \pi G d^2 / \dot E$ \dotfill & $0.10_{-0.04}^{+0.05}$ $(<0.13)$ \\
\enddata
\tablecomments{See Section \ref{sec:gammaanalysis} for details on the measurement of these parameters. Quoted errors are statistical, while numbers in parentheses represent canonical values, obtained by fixing the photon index of \psr{} at 1.3 in the spectral analysis.}
\end{deluxetable}

As illustrated in \citet{Kerr2011weights} and \citet{Guillemot2012a}, weighting the \g-ray events by the probability that they originate from the putative pulsar increases the sensitivity to faint pulsations. To measure these probabilities we performed a binned likelihood analysis of the spectra of the sources in the region of interest, using the \texttt{PyLikelihood} module of the STs. Our spectral model included the 50 2FGL catalog sources \citep{Fermi2FGL} within 20$^\circ$ of \psr{}. The contribution from the pulsar, which is not associated with any 2FGL source, was modeled as an exponentially cutoff power law (ECPL) of the form $dN/dE \propto E^{-\Gamma} \exp \left( - E / E_c \right)$, where $\Gamma$ is the photon index and $E_c$ is the cutoff energy of the spectrum. The source model also included contributions from the extragalactic diffuse emission and the residual instrumental background, jointly modeled using the \emph{iso\_p7v6source} template, and from the Galactic diffuse emission, modeled with the \emph{gal\_2yearp7v6\_v0} map cube\footnote{See http://fermi.gsfc.nasa.gov/ssc/data/access/lat/BackgroundModels.html}. The parameters of \psr{}, of the seven sources within 10$^\circ$ of the pulsar, and the normalization factors of diffuse components were left free in the fit. A Test Statistic \citep[TS; for a definition see][]{Fermi2FGL} value of 20.3 was found for \psr{} at this stage.

To improve the quality of the spectral results we analyzed the $[0 ; 0.2] \cup [0.5 ; 0.8]$ phase interval (OFF pulse), removing the contribution from the pulsar. This allowed us to obtain an improved fit of the spectra of neighboring sources. Sources with TS values below 2 were excluded from the best-fit model obtained at this point, and the spectral parameters of sources beyond 3$^\circ$ from the double pulsar system were frozen at their best-fit values. We finally analyzed events in the complementary $[0.2 ; 0.5] \cup [0.8 ; 1]$ phase interval (ON pulse) with the resulting source model, refitting the contributions of the pulsar and of the three sources within 3$^\circ$, to determine the spectral parameters for \psr{} listed in Table \ref{tab:params}. A TS value of 31.2 is obtained for \psr{}. Refitting the data with a power law shape for the pulsar, we found that the ECPL model is preferred by the likelihood at the \cutoffsigma{}$\sigma$ level. A cross-check of this analysis performed with the independent \textsc{Pointlike} analysis tool \citep{KerrThesis} yielded results consistent with those listed in Table \ref{tab:params}. Also listed in the table are the \g-ray luminosity $L_\gamma = 4 \pi f_\Omega G d^2$ where $G$ is the phase-averaged energy flux above 0.1 GeV and assuming a beaming correction factor $f_\Omega = 1$, and the derived efficiency $\eta = L_\gamma / \dot E$. The photon flux $F$ and energy flux $G$ of \psr{} are among the lowest values of any \g-ray pulsars detected to date, probably a consequence of its low $\dot E$ value. We also note that \psr{} has the lowest value of the magnetic field at the light cylinder of any recycled pulsars detected in \g{} rays, $B_{\rm LC} \sim 4.97 \times 10^3$ G \citep[see Equation 3.22 of][]{Handbook}.

\begin{figure}
\epsscale{1.15}
\centerline{
\hfill
\plotone{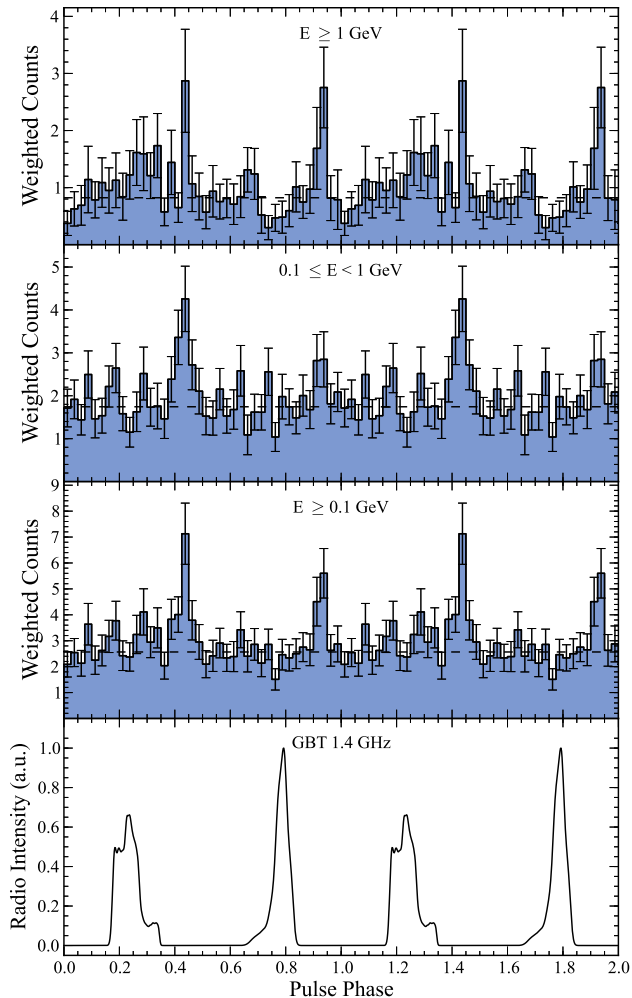}
\hfill
}
\caption{Radio and \g-ray light curves for \psr{}. Bottom panel: 1.4 GHz radio profile based on observations made with the Green Bank Telescope. Upper panels: \emph{Fermi} LAT profiles in different energy bands, obtained by weighting each event by its probability to have been emitted by \psr{}. Photons with probabilities smaller than 0.05 were excluded. Horizontal dashed lines indicate the estimated background levels (see \citet{Guillemot2012a} for the determination of these lines). The \g-ray pulse profiles have 40 bins per rotation. Two rotations are shown for clarity. \label{fig:lc}}
\end{figure}

None of the different attempts to model the spectrum of \psr{} presented above yielded satisfactory measurements of its photon index, $\Gamma$. In all cases, a small value consistent with 0 was favored. This is likely a consequence of the cutoff energy $E_c$ being too low to reliably determine the photon index $\Gamma$. Limits on the other spectral parameters can thus be placed by repeating the analysis using a value for the photon index which is representative of the $\Gamma$ values observed for other MSPs. Table~\ref{tab:params} lists the results obtained by fixing $\Gamma$ at 1.3, this value being the average photon index measured for the strongest MSPs in 2PC. The spectral parameters quoted in parentheses represent canonical values, arising from the lack of constraints on $\Gamma$.

The tool \texttt{gtsrcprob} was finally used to assign each event a probability that it originated from \psr{} based on the fluxes and spectra obtained from the likelihood analysis. The integrated pulse profile of \psr{} over 0.1 GeV and for probabilities larger than 0.05 is shown in Figure \ref{fig:lc}. The zero of phase is defined by the maximum of the first Fourier harmonic of the signal transferred back to the time domain. We find a weighted $H$-test parameter of \whts{} \citep{Kerr2011weights,deJager2010}, corresponding to a significance of $\sim$\whtssigma{}$\sigma$.

As can be seen from Figure \ref{fig:lc}, the \g-ray profile comprises two peaks, with indications for additional complexity. We fitted the profile above 0.1 GeV with Lorentzian functions and found the peak positions and full widths at half-maxima for the first and second \g-ray peaks as listed in Table \ref{tab:params}. Defining $\delta_1$ and $\delta_2$ as the separation between the \g-ray peaks and the maxima of the closest 1.4 GHz radio peaks, respectively at $\Phi_{r,1} \sim 0.24$ and $\Phi_{r,2} \sim 0.79$, we find $\delta_1 = 0.20 \pm 0.01$ and $\delta_2 = 0.13 \pm 0.01$. The uncertainty on $\delta_1$ and $\delta_2$ caused by the error on the dispersion delay is $\Delta (\mathrm{DM}) / (k f^2)$, where $\Delta (\mathrm{DM}) = 5 \times 10^{-3}$ pc cm$^{-3}$ is the uncertainty on the dispersion measure (DM) reported in \citet{Kramer2006}, $k$ is the dispersion constant \citep[see][]{Handbook} and $f = 1.4$ GHz, is found to be $\sim 5 \times 10^{-4}$ in phase, and can thus be neglected. The \g-ray emission from \psr{} is thus offset from the radio emission, suggesting distinct origins in the magnetosphere of the pulsar. This also holds for the X-ray peaks detected by \citet{Chatterjee2007} in \emph{Chandra} HRC data, which are likewise aligned with the radio peaks.

Finally, we have searched the OFF-pulse fraction of the data for modulation at the orbital period, caused by the collision of the particle winds from the pulsars, and we have also searched the ON-pulse signal for attenuation of the emission from pulsar A around conjunction, caused by, e.g., photon--light-pseudoscalar-boson oscillation in the magnetosphere of B, as proposed by \citet{Dupays2005}. In both cases, we observed only steady emission as a function of orbital phase.

\section{Constraining the Viewing \\ Geometry of \psr}
\label{sec:geom}

We can place constraints on the magnetic inclination $\alpha$ and viewing angle $\zeta$ by modeling the radio and \g-ray light curves and the radio polarization. For all models we assume the vacuum, retarded-dipole magnetic field geometry \citep{Deutsch1955}. The emission is assumed to originate in the open zone, determined by field lines that do not close within the light cylinder (at $R_{\rm LC} = c / \Omega = c P / 2 \pi$), traced to foot points on the star surface at $R_{\rm NS}$ (radius of the neutron star), which define the polar cap. The computation and fitting procedure are described in \citet{Watters2009}, \citet{Venter2009}, \citet{RomaniWatters2010}, and references therein.

In two of the prevalent models of \g-ray emission, uniformly emissive zones of the magnetosphere stand in for more physical models. The first is the outer gap picture \citep[OG;][]{Cheng1986,Yadigaroglu1995} in which radiation is emitted between the null charge surface where $\mathbf{\Omega} \cdot \mathbf{\mathit{B}} = 0$ and $R_{\rm LC}$. The second is the two-pole caustic model \citep[TPC;][]{Dyks2003}, which we take to be a geometric realization of the slot gap model \citep{Muslimov2004}. In the original TPC model emission extends from the surface to 0.75 $R_{\rm LC}$. In both geometries the \g-ray emission is confined toward the edge of the open zone coinciding with the last open field lines, in a ``gap'' idealized as a region of width $w_{\rm em}$, interior to an accelerating layer of width $w_{\rm acc}$. In the OG model the \g-ray emission originates from a thin layer on the inner edge of the gap, closest to the magnetic pole, whereas in the TPC model emission is produced throughout the gap. The radio emission can occupy a large fraction of the open field lines, albeit generally at lower altitude.

\subsection{Joint Radio and \g-Ray Light Curve Fits}
\label{sec:lcmodeling}

One approach is to fit the \g-ray and radio light curves directly, following the geometrical light curve modeling of \citet{Venter2009}, with the fitting procedure described in more detail in, e.g., \citet{Guillemot2012a}. For these computations we assume $P = 25$ ms and $\dot{P}=1\times10^{-18}$ s s$^{-1}$, and treat the radio pulse as a classical single-altitude, hollow-cone beam centered on the magnetic axis as described by \citet{Story2007}. For this analysis we start from a grid of models with 1$^\circ$ steps in both $\alpha$ and $\zeta$, and fit the emission zone accelerating and emitting gap widths, $w_{\rm acc}$ and $w_{\rm em}$, with steps of 2.5\% of the polar cap opening angle ($\Theta_{\rm PC} \approx (\Omega R_{\rm NS}/c)^{1/2}$). For this computation we extend the emission to 0.95 $R_{\rm LC}$.

The models are fit to an unweighted-counts \g-ray light curve for \psr{}, constructed by selecting events from the data described in Section \ref{sec:gammaanalysis} found within 2$^\circ$ of the pulsar, with energies above 0.1 GeV and with probabilities of being associated with the pulsar larger than 0.05. In the case of low-statistics \g-ray light curves with sharp peaks such as those observed for \psr{}, using a $\chi^2$ statistic with binned light curves may not be optimal as this statistic is often insensitive to the peaks. We have thus found that it is preferable to use the unweighted counts, which allow the use of Poisson likelihood, as opposed to the weighted counts which require using a $\chi^2$ statistic for binned light curves. The radio profile is fit with a $\chi^{2}$ statistic, assuming the same relative uncertainty for each phase bin, which is combined with the likelihood from the \g-ray fit to determine the best-fit model parameters from a scan over the model phase space. We used 30 bins for both the \g-ray and radio light curves in these fits.

One difficulty in joint fitting of the radio and \g-ray profiles is that the low statistical uncertainty of the radio light curve compared to the uncertainty of the \g-ray bins causes the fit to be dominated by the radio data and to ignore the constraints provided by the \g{} rays. Because our radio model is simplistic, this can lead to unrealistic solutions. We have therefore investigated different prescriptions for the ``uncertainty'' on the radio profile bins, in order to balance the radio and \g-ray contributions to the joint likelihood. The following prescription gives the most satisfactory results. First, we select an on-peak interval for the \g-ray light curve, which should dominate the likelihood, and define $\sigma_{\rm g\,ave}$ as the average, fractional uncertainty of the \g-ray bins in this interval. We then define a radio uncertainty $\sigma_{\rm r} = r_{\rm max}\times\sigma_{\rm g\,ave}$, where $r_{\rm max}$ denotes the maximum radio light curve bin. Although this prescription gives satisfactory results, it is arbitrary. We tested its sensitivity by varying $\sigma_{\rm r}$ by a factor of two \citep[see][]{Guillemot2012b}; the resulting best-fit geometries changed by $\alpha \lesssim 4^{\circ}\ (12^{\circ})$ and $\zeta \lesssim 4^{\circ}\ (9^{\circ})$ for the TPC (OG) model. The latter values can be considered systematic errors on our fit results.

\begin{figure*}
\epsscale{0.85}
\centerline{
\hfill
\plotone{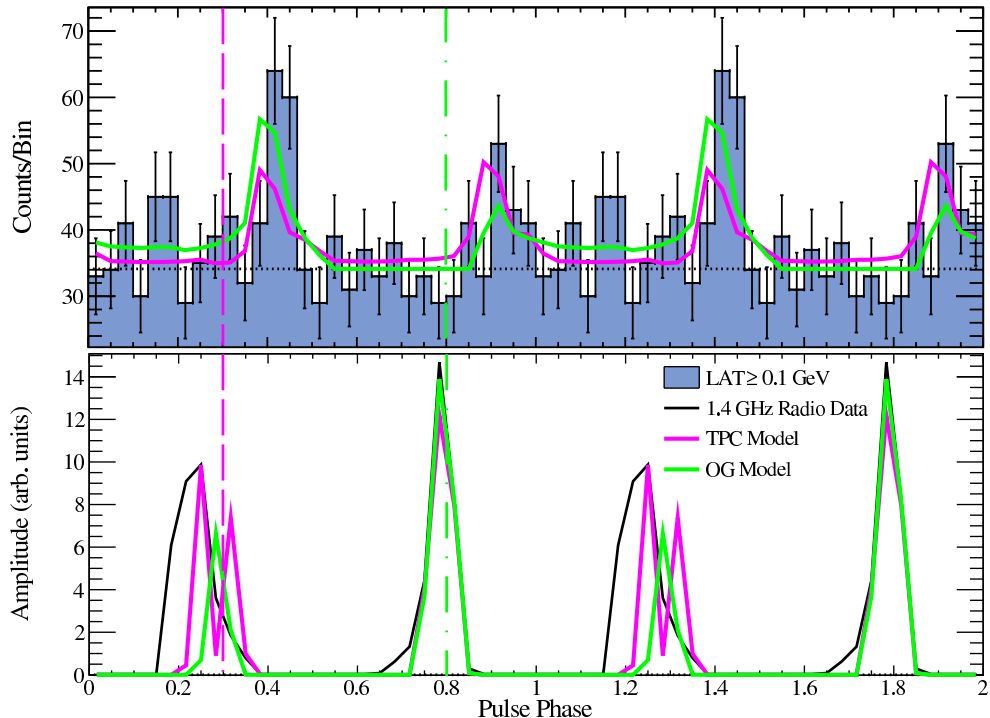}
\hfill
}
\caption{Top: modeled light curves and \g-ray data for \psr{}. Bottom: 1.4 GHz radio profile and best-fitting profiles. TPC light curves are shown as pink lines and OG light curves are shown as green lines. The vertical dashed pink line and dash-dotted green line indicate the closest approach to the magnetic axis under the best-fit TPC and OG models, respectively. \label{fig:lcmodeling}} 
\end{figure*}

The best-fitting model light curves are shown in Figure~\ref{fig:lcmodeling}, and the parameters are given in Table~\ref{tab:fitpars}. The $1\sigma$ uncertainties are estimated from two-dimensional likelihood profiles for $\alpha$ and $\zeta$ and one-dimensional profiles for the other parameters. The best-fit gap widths for both models have size 0, indicating that they are unresolved by our simulations. The simple geometric models we use neglect important physics, e.g., the effects of magnetospheric charges and currents on the gaps, and cannot perfectly fit the data. We use this discrepancy to set the scale of our parameter constraints by rescaling the likelihood by its best fit value, reduced by one-half the degrees of freedom (dof). In the Gaussian approximation, this is equivalent to setting the best-fit reduced $\chi^2=1$. With this normalization, the $\chi^{2}$ differences between the TPC and OG models are not significant. Also shown in Figure~\ref{fig:lcmodeling} are the closest approaches to the magnetic axis under the best-fit TPC and OG light curves. We note that our modeling finds different locations for the magnetic axis under the two geometries, due to the ambiguity in defining the first and second \g-ray peaks when the phase separation is close to 0.5.

Using these geometric models, the light curves obtained when reflecting across $\alpha = 90^\circ$ and $\zeta = 90^\circ$ are the same, but shifted by 0.5 in phase. We report confidence regions for $\alpha < 90^\circ$ and $\zeta < 90^\circ$, but note that identical, reflected regions exist in the other three quadrants.

From the simulations, we can estimate the beaming correction factor $f_{\Omega}$ \citep{Watters2009} which relates emission along a given line-of-sight to the total flux. The values for the models are given in Table~\ref{tab:fitpars} with estimated uncertainties based on the confidence regions in the $\alpha$--$\zeta$ plane. Both models suggest that this factor should be $\sim$1, implying corrected \g-ray efficiencies close to 10\%.

\begin{deluxetable*}{l c c c c c c c c}
\tablecaption{Radio and \g-Ray Light Curve Fit Parameters for \psr{}\label{tab:fitpars}}
\tablecolumns{9}
\tablehead{\colhead{Model} & \colhead{$\alpha$ $(^{\circ})$} & \colhead{$\zeta$ $(^{\circ})$} & \colhead{$w_{\rm acc}$\tablenotemark{a} (\% $\Theta_{\rm PC})$} & \colhead{$w_{\rm em}$\ (\% $\Theta_{\rm PC})$} & \colhead{$-\ln(\rm likelihood)$} & \colhead{dof} & \colhead{$f_{\Omega}$}}
\startdata
TPC & $80^{+9}_{-3}$ & $86^{+2}_{-14}$ & . . . & 0.0$\pm$2.5 & 121.0 & 54 & $0.89^{+0.15}_{-0.20}$ \\
OG & $88^{+1}_{-17}$ & $74^{+14}_{-4}$ & 0.0$\pm$10.0 & 0.0$\pm$2.5 & 123.1 & 53 & $0.95^{+0.10}_{-0.37}$
\enddata
\tablenotetext{a}{In the TPC model the accelerating and emitting gaps are the same.}
\tablecomments{Using these geometric models the results are invariant under either a reflection across $\alpha = 90^\circ$ or $\zeta = 90^\circ$. }
\end{deluxetable*}

In addition to standard TPC and OG models, we have explored two alternative geometries. In the first approach, a very low-altitude radio cone (low-altitude slot gap radio geometry) was invoked in conjunction with the usual TPC model for the \g{} rays. This was done in the context of a radio cone producing peaks leading the caustic \g-ray peaks. The second alternative, motivated by the idea that the radio may indeed have a dominating leading peak with the radio profile lagging the \g-ray light curve, assumed conal radio \emph{and} \g-ray geometries, with the \g-ray cone being \emph{lower} than that of the radio. The fits from these alternative approaches were not satisfactory, as they were unable to reproduce the observed radio and \g-ray peak shapes and separations, leading us to abandon these scenarios. However, they point to interesting avenues of model refinement, e.g., investigation of cones with non-uniform emissivities (such as patchy or one-sided radio cones), or non-aligned \g-ray and radio cones. Such refinements, which are expected to lead to improved light curve fits, are beyond the scope of the current paper and will be developed in future work \citep{Seyffert2013}.

We remark that the fitting results presented in this section were obtained by assuming that the observed radio emission from \psr{} originates from both magnetic poles. It has been proposed that the non-$180^\circ$ separation of the two radio peaks and the overall symmetry of the profile imply that the radio emission originates from magnetic field lines associated with a single pole \citep[see, e.g.,][]{Manchester2005}. However, \citet{Ferdman2013} showed that this scenario would imply a high altitude origin for the radio emission in the pulsar's magnetosphere. If both the radio and the \g{} rays originate from the outer magnetosphere, we would expect the pulses to be aligned in phase \citep{Venter2012}. The two-pole origin for the radio emission therefore seems to be a more natural solution.

\subsection{Polarization Fitting}
\label{sec:pol}

When attempting to understand pulsar viewing geometry, radio polarization can provide very powerful constraints. Although recycled pulsars are notoriously difficult to model, \psr{} shows substantial linear polarization structure, and thus offers good prospects of constraining the orientation. In our polarization modeling we follow the conventional assumption that the electric vector position angle (P.A.) follows the projection onto the plane of the sky of the magnetic field line at the emission point. The P.A. sweep as one moves past the radio pole(s) thus probes the viewing geometry, with particularly strong constraints on $\beta=\zeta-\alpha$. Our modeling extends beyond the point dipole rotating vector model \citep[RVM;][]{Radhakrishnan1969}, to follow the distortion of the magnetic field as the emission zone moves to a non-negligible fraction of $R_{\rm LC}$ \citep{Craig2012}. In addition, we follow \citet{Karastergiou2009} in treating ``orthogonal mode jumps'' for which the P.A. shifts by $\pm 90^\circ$, and optionally account for the effects of interstellar scattering (negligible for \psr{}). For details see \citet{Craig2013}.

\begin{deluxetable}{c c c c c c c c c}
\tablecaption{Polarization Fit Parameters for \psr{}\label{tab:polarfitpars}}
\tablecolumns{7}
\tablehead{\colhead{$\alpha_f$ $(^{\circ})$} & \colhead{$\zeta_f$ $(^{\circ})$} & \colhead{$r_{1}/R_{\rm LC}$} & \colhead{$r_{2}/R_{\rm LC}$} & \colhead{$\chi^{2}$} & \colhead{dof} & \colhead{$\Delta\phi$}}
\startdata
${98.8}^{+8}_{-1.5}$ & ${95.8}^{+13.2}_{-4.3}$ & $0.01^{+0.22}_{-0.01}$ & $0.11^{+0.49}_{-0.05}$ & $48$ & $35$ & $0.443^{+0.008}_{-0.055}$
\enddata
\tablecomments{Errors are the extrema of the $1\sigma$ contours in the full multidimensional parameter space. $r_{1}$ is the emission altitude of the central component of P1 and $r_{2}$  the altitude of the central component of P2. $\Delta\phi$ marks the offset of the closest approach of the surface magnetic axis (in P1) from the total intensity peak in P2; this is phase 0.243 in Figure~\ref{fig:lc}.}
\end{deluxetable}

\begin{figure*}
\epsscale{0.75}
\centerline{
\hfill
\plotone{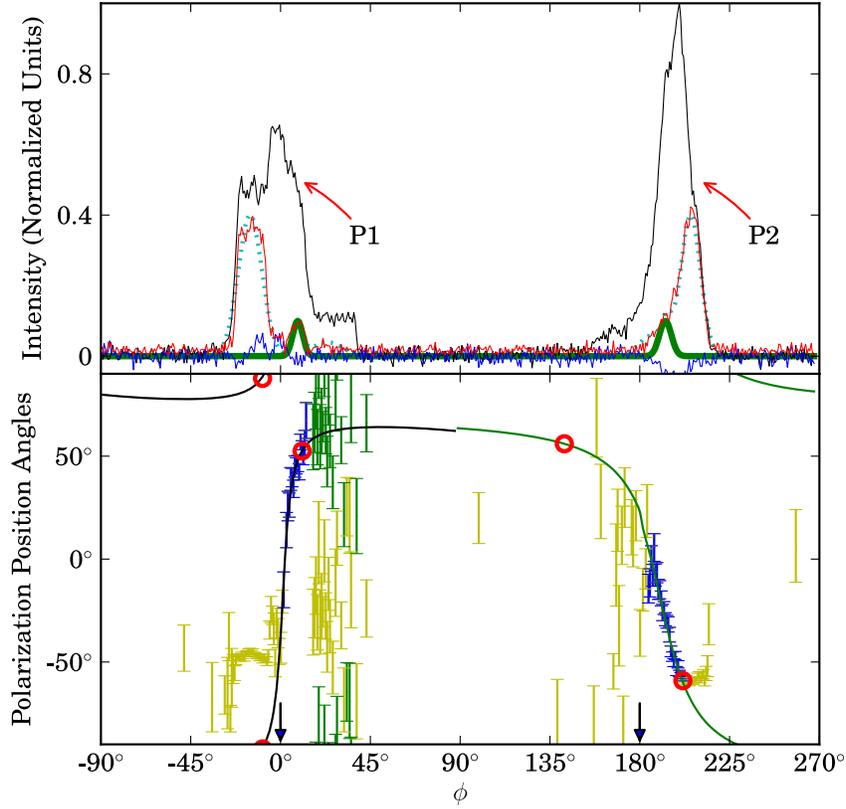}
\hfill
}
\caption{Polarimetric profile for \psr{} as observed at 1.4 GHz with the Parkes radio telescope (R.~N.~Manchester, private communication). Top: Stokes parameter curves (black: I, red: L, blue: V) and Gaussian decomposition of the linear intensity components (dotted curve is all linear, the thick green curve gives the rapid sweep central component fit here). Bottom: position angle data (blue: central components, yellow: other P.A. values, green: P1 tail with an orthogonal mode jump). The smooth curves give the best fit model for the two poles while the red circles denote the boundaries of the open zone at the emission altitude. Arrows denote the phase of the closest approach of the magnetic axes to the Earth line-of-sight. The smallest impact parameter $\beta$ is for P1 at phase $\phi_B=0$ (phase of the magnetic dipole).\label{fig:polarfit}} 
\end{figure*}

\begin{figure}
\centerline{
\hfill
\plotone{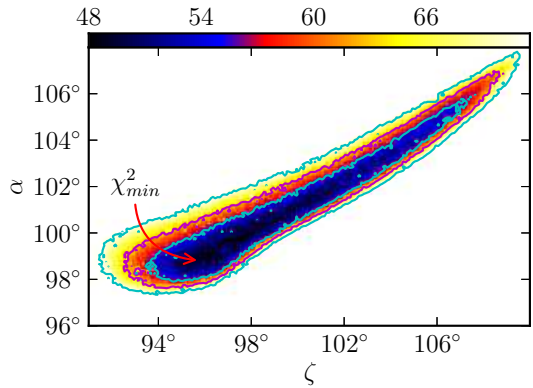}
\hfill
}
\caption{Projection of the $\chi^2$ surface of the RVM fit to the radio polarization data (central components) for \psr{} onto the $\alpha$--$\zeta$ plane. The best fit orientation angles ($\alpha,\, \zeta$) are indicated by an arrow, while the contours are shown at the $1\sigma$, $2\sigma$ and $3\sigma$ levels.\label{fig:polarchisq}} 
\end{figure}

Figure~\ref{fig:polarfit} shows a polarimetric profile for \psr{} measured at 1.4 GHz with the Parkes radio telescope (R.~N. Manchester, private communication). The Stokes parameters and Gaussian decomposition of the linear intensity components are displayed in the top panel, while the lower panel shows the P.A. data. In this section we refer to the brightest radio peak as P2. Both it and P1 are multi-component. In polarization fitting, these components are interpreted as different emission zones with different linear and circular polarization fractions and, possibly, different orthogonal mode states and emission altitudes.  Progressively across the P1 pulse we see a component with strong polarization at nearly constant P.A., a largely unpolarized peak, a component with rapid linear polarization sweep after the maximum and a separate, weakly linear polarized component on the flat pulse tail. For P2 the pattern appears reversed, with a weak linear tail leading the pulse, a linear component with rapid P.A. sweep, an unpolarized peak and then a strongly linearly polarized component with little P.A. sweep. The flat P.A. components at the front of P1 and the back of P2 are not naturally produced by any component locked to a local magnetic field. We suspect that these components are controlled by magnetospheric plasma, either in fixing the P.A. directly or in strongly distorting the field at very high altitude. Here we fit the rapid P.A. sweep in the central linearly polarized component of each pulse. The parameters are $\alpha$, $\zeta$, $\phi_0$ (the phase of the total intensity peak with respect to the surface dipole magnetic axis), $\Psi_0$, the P.A. offset and the two emission heights $r_1$ and $r_2$ in units of $R_{\rm LC}$. The fit values, the $\chi^{2}$ and the full projected (multi-parameter) uncertainty ranges on the fit parameters are shown in Table \ref{tab:polarfitpars}. Figure~\ref{fig:polarchisq} shows a projection of the $\chi^2$ surface of the fit onto the $\alpha$--$\zeta$ plane.

The best-fit P.A. models for the two poles are displayed in the lower panel of Figure~\ref{fig:polarfit}. Note that the weak trailing linear component in P1 is separated from the central component by a zero in the linear intensity L, and so is very likely an orthogonal mode jump. The jumped P.A. points cluster around the model prediction. We do not attempt to model the total intensity light curve or the P.A. behavior of the flat P.A. components leading P1 and trailing P2, since these do not follow simple magnetospheric models. However, we can make some comments about the origin of this emission. First, the circles mark the phase ranges of the open zone at the altitude fit for the two central components. These components occupy the second half of the open zone at this altitude, terminating at the closed zone boundary, an excellent consistency check. However, the wings of the pulses, especially at large $\phi$, lie beyond the circles and hence must either come from higher altitude or else arise from the classical closed zone. Our picture of each pole is thus a cut through a hollow cone with the central component arising from low altitude and higher altitude components from the pulse wings. Since P1 has the smallest $|\beta|=3^\circ$, it is viewed at lower altitude than P2 with $\beta=14\fdg6$.

These fits also determine the phase of closest approach of the surface dipole axis. This can be compared with predictions of the outer magnetosphere \g-ray models at the fit $\alpha$ and $\zeta$. We follow \citet{RomaniWatters2010} in examining the $\alpha$ and $\zeta$ for the best fits to the \g-ray light curve and then checking against polarization constraints. For the OG model the best light curve matches are found very close to the angles fit to the polarization data.  For the TPC model (here limited to $0.75 R_{\rm LC}$), the best fits also occur near $(90^\circ,\, 90^\circ)$. These are somewhat worse than the best OG values, since they include too much unpulsed emission at all phases, but they are quite acceptable.  In this regard the polarization fitting agrees well with the fits of Section \ref{sec:lcmodeling}, when transformed to the $\alpha>90^{\circ}$ and $\zeta>90^{\circ}$ region. However, the very accurate phase measurements from the P.A. sweep highlight a tension between the two bands: the radio-determined pole is always $\sim 0.1$ in phase before the best-fit axis for the \g{} pulse. This tension is also evident in the fits to the cruder pulse profiles of Section \ref{sec:lcmodeling}. However, non-vacuum effects provide a plausible explanation for this offset. In \citet{Kalapotharakos2012} it was noted that the primary effect on the high altitude pulse of increasing the magnetosphere conductivity was a growing lag in phase with respect to the surface dipole. Lags of $\delta \phi \sim 0.1$ were seen for high conductivity magnetospheres.  If we disconnect the \g-ray and radio zero-phase definitions in the joint light curve fits, letting $\delta$ be larger by as much as 0.1 in phase, the best-fit $\alpha$ and $\zeta$ values changed by $\leq4^{\circ}$, generally making the agreement between the two methods better.

\begin{figure*}
\epsscale{0.75}
\centerline{
\hfill
\plotone{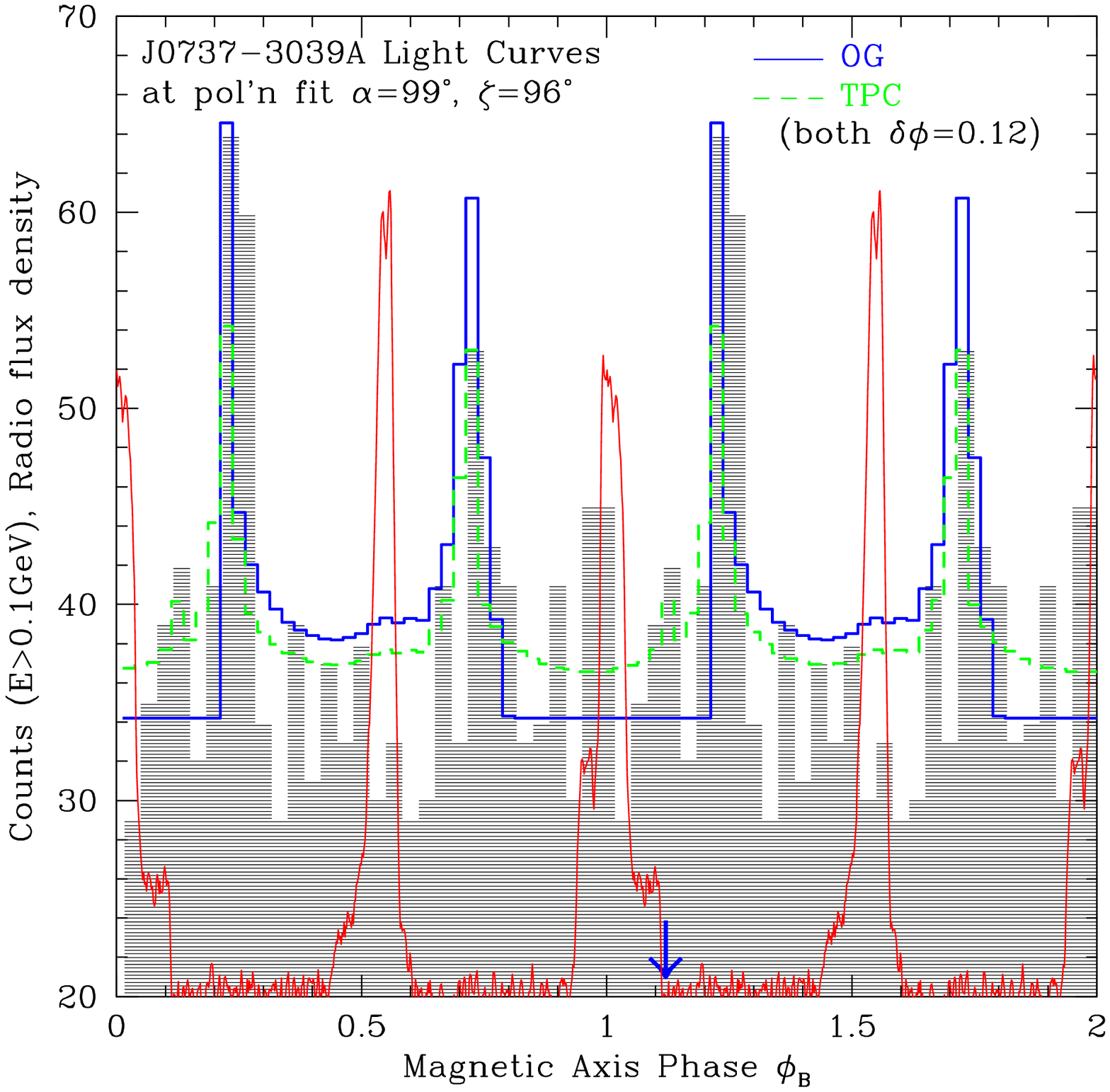}
\hfill
}
\caption{Light curves for \psr{} at the $\alpha$ and $\zeta$ angles determined from the radio polarization study, for the OG model (solid blue line) and the TPC model (dashed green line). The observed radio and \g-ray profiles are shown as a solid red line and as a shaded gray histogram, respectively. See Figure~\ref{fig:polarfit} for the definition of the magnetic axis phase $\phi_B$. The blue arrow denotes the location of the magnetic axis under the OG and TPC geometries. \label{fig:lcmodeling2}}
\end{figure*}

In Figure~\ref{fig:lcmodeling2} we show the predicted OG and TPC \g-ray light curves at the pulsar geometry determined from the polarization study, superimposed on the radio and the \g-ray data, and using the phase of closest approach to the surface dipole axis as defined by the polarization fits. Both modeled light curves are acceptable, with the OG one being somewhat better (reduced $\chi^2$ of 1.45 compared to 1.56 for the TPC model). For these $\alpha$ and $\zeta$ angles we find beaming correction factors $f_\Omega$ of 0.92 for the OG model and 1.15 for the TPC model, similar to the ones given in Table~\ref{tab:fitpars} and therefore also implying \g-ray efficiencies of $\sim 10\%$.

\section{Discussion}
\label{sec:discussion}

\subsection{The Low Spectral Cutoff Energy and Magnetic Field at the Light Cylinder of \psr{}}

In all outer-magnetospheric models, we expect the accelerating electric field $E_{||}$ to scale as \citep[e.g.,][]{FermiVela2}:

\begin{eqnarray}
E_{||} \propto C(r) B_{\rm LC} w_{\rm acc}^2,
\end{eqnarray}

with $C(r)$ some function of radius, and $w_{\rm acc}$ the accelerating gap width. On the other hand, if leptons are accelerated in the radiation-reaction-limited regime, i.e., when the gain in energy due to acceleration by the electric field is balanced by curvature radiation losses, we expect that the spectral cutoff energy should scale as:

\begin{eqnarray}
E_{c} \propto E_{||}^{3/4} \rho_c^{1/2} \propto C(r)^{3/4} w_{\rm acc}^{3/2} B_{\rm LC}^{3/4} \rho_c^{1/2},
\end{eqnarray}

with $\rho_c$ the radius of curvature (at the position where the photons of maximal energy are produced). If we take $C(r)$, $\rho_c$ and $w_{\rm acc}$ constant for all MSPs (e.g., independent of $P$ and $\dot{P}$), this leads us to expect: 
 
\begin{eqnarray}
E_{c} \propto B_{\rm LC}^{3/4}.
\end{eqnarray}

One does see a trend when plotting published values of $E_{c}$ versus $B_{\rm LC}$ for all known gamma-ray MSPs, but it is much weaker than expected, and the index is much smaller than 0.75, implying that our assumption of constant $C(r)$, $\rho_c$ and $w_{\rm acc}$ across the MSP population is too simplistic. \psr{} is a mildly recycled MSP, having a relatively long $P$ and hence the lowest $B_{\rm LC} \propto P^{-5/2}\dot{P}^{1/2}$ of all currently known gamma-ray MSPs. Assuming that the above empirical trend may be extrapolated down to low $B_{\rm LC}$ values, one would expect a relatively low cutoff energy. This is indeed seen, although \psr{}'s cutoff energy falls significantly below even the extrapolated trend.

Apart from invoking the weak empirical trend discussed above, one may more generally observe that the relatively long period of \psr{} implies a lower acceleration potential and hence a lower spectral cutoff than for typical MSPs, as is observed from the spectral analysis.

\subsection{Constraints on the System's Formation History}

From the modeling of the radio and \g-ray light curves of \psr{} under the OG and the TPC geometries, and the modeling of radio polarization data (see Section~\ref{sec:geom}), we constrained the magnetic inclination angle $\alpha$ and the viewing angle $\zeta$ to be close to 90$^\circ$. The best-fit $\alpha$ and $\zeta$ angles obtained from the polarization study (see Table~\ref{tab:polarfitpars}) agree well with those obtained from the light curve modeling (Table~\ref{tab:fitpars}). The conclusion that $\alpha$ and $\zeta$ are close to $90^\circ$ is in line with the results of \citet{Ferdman2013}, based on the analysis of six years of radio observations of \psr{}, revealing no significant variations of its radio profile with time. \citet{Ferdman2013} find $\alpha$ and $\zeta$ values of $90\fdg2^{+16.3}_{-16.2}$ and $90\fdg8^{+0.27}_{-0.46}$, respectively. These values were calculated assuming that radio emission is seen from both magnetic poles, that the orbital inclination is $88\fdg7$, and averaging the results from all radio pulse heights used (30\% to 50\%) and all observation epochs.

By virtue of the relationship between $\zeta$ and the misalignment angle between the spin axis and the orbital angular momentum, $\delta_{SO}$ \citep[cf. Equation (3.36a) of][]{Damour1992}, our values for $\zeta$ close to $90^\circ$ imply small $\delta_{SO}$ values, unless the pulsar is at a special precession phase. Nevertheless, the fact that \citet{Ferdman2013} find a 95\% confidence upper limit on $\delta_{SO}$ of $\sim 3\fdg2$ from their observations taken between 2005 June and 2011 June indicates that the pulsar was very unlikely to be at a special precession phase during the \emph{Fermi} observation. Our constraints on $\zeta$ therefore provide an independent confirmation that $\delta_{SO}$ is close to 0$^\circ$. Since the spin axis of pulsar A and the angular momentum of the binary are likely to have become aligned through the accretion of matter from pulsar B's progenitor star by pulsar A, a value close to 0$^\circ$ for the misalignment angle $\delta_{SO}$ supports the scenario under which a small kick was imparted to the system by the supernova explosion of pulsar B's progenitor star. An electron-capture supernova, resulting from a collapsed O-Ne-Mg core as proposed by \citet{Podsiadlowski2005}, could provide such a formation scenario for pulsar B \citep[see][and references therein for a detailed discussion]{Ferdman2013}.

Because \psr{} is a faint \g-ray emitter, we were not able to investigate any evolution of the \g-ray profile as a function of time caused by a variation of the line-of-sight angle, $\zeta$, in the four years of LAT data considered here. Since the misalignment angle $\delta_{SO}$ has been shown to be close to 0$^\circ$, changes in $\zeta$ are likely small, but are still expected. An exciting prospect, when significantly more \emph{Fermi} LAT data are available, will be to monitor the evolution of the \g-ray profile of \psr{}. To date, no \g-ray pulsar light curves have even been observed to vary with time. 

\acknowledgements

We are very grateful to Richard N. Manchester (ATNF) for providing valuable comments which helped to improve the manuscript, for providing us with the Parkes polarization data analyzed in this paper, and for carrying out independent cross-checks of the relative alignment of the radio and \g-ray profiles using Parkes timing data taken concurrently with the \emph{Fermi} dataset. The Parkes radio telescope is part of the Australia Telescope, which is funded by the Commonwealth of Australia for operation as a National Facility managed by the Commonwealth Scientific and Industrial Research Organisation (CSIRO). 

We also wish to express our gratitude to Isma\"el Cognard (LPC2E, Orl\'eans) for providing Nan\c{c}ay timing solutions for PSR~J0737$-$3039A, that were used for searching for pulsations before this manuscript was written. The Nan\c{c}ay Radio Observatory is operated by the Paris Observatory, associated with the French Centre National de la Recherche Scientifique. 

The \textit{Fermi} LAT Collaboration acknowledges support from a number of agencies and institutes for both development and the operation of the LAT as well as scientific data analysis. These include NASA and DOE in the United States, CEA/Irfu and IN2P3/CNRS in France, ASI and INFN in Italy, MEXT, KEK, and JAXA in Japan, and the K.~A.~Wallenberg Foundation, the Swedish Research Council and the National Space Board in Sweden. Additional support from INAF in Italy and CNES in France for science analysis during the operations phase is also gratefully acknowledged. Portions of this research performed at NRL are sponsored by NASA DPR S-15633-Y.

\bibliographystyle{apj}

\bibliography{J0737-3039A} 

\end{document}